\documentclass[journal]{IEEEtran}
\usepackage{tabularx}

 \usepackage{amsmath,amstext,amsfonts,amssymb,mathrsfs}
\usepackage{algorithm,algorithmic,tikz}
\usepackage{hyperref}
\usepackage[sort,compress]{cite}
\usepackage{lipsum}
\usepackage{caption}
\usepackage{subcaption}
\usepackage{comment}
\usepackage{multirow}

\makeatletter
\newcommand\fs@betterruled{%
  \def\@fs@cfont{\bfseries}\let\@fs@capt\floatc@ruled
  \def\@fs@pre{\vspace*{5pt}\hrule height.8pt depth0pt \kern2pt}%
  \def\@fs@post{\kern2pt\hrule\relax}%
  \def\@fs@mid{\kern2pt\hrule\kern2pt}%
  \let\@fs@iftopcapt\iftrue}
\floatstyle{betterruled}
\restylefloat{algorithm}
\makeatother
\usepackage{bbm}

\title{\Huge Dynamic Matching Markets in Power Systems: Concepts and Solution using Deep Reinforcement Learning}

\author{Majid Majidi$^{*}$,~\IEEEmembership{Student Member,~IEEE}, Deepan Muthirayan$^{*}$,~\IEEEmembership{Member,~IEEE}, Masood Parvania,~\IEEEmembership{Senior Member,~IEEE}, Pramod P. Khargonekar,~\IEEEmembership{Fellow,~IEEE}
\thanks{This work is supported in part by the National Science Foundation under Grant ECCS-1839429.}
\thanks{M. Majidi and M. Parvania are with the Department of Electrical and Computer Engineering, The University of Utah (emails: majid.majidi@utah.edu, masood.parvania@utah.edu).}
\thanks{D. Muthirayan and P. P. Khargonekar are with the Department of Electrical Engineering and Computer Sciences, University of California Irvine, Irvine (emails: deepan.m@uci.edu, pramod.khargonekar@uci.edu). $^*$Both authors made equal contribution.}
}
\begin{document}

\maketitle

\begin{abstract}
Traditional bulk load flexibility options, such as load shifting and load curtailment, for managing uncertainty in power markets limit the diversity of options and ignore the preferences of the individual loads, thus reducing efficiency and welfare. 
This paper proposes an alternative to bulk load flexibility options for managing uncertainty in power markets: a reinforcement learning based dynamic matching market.  
We propose a novel hybrid learning-based model for maximizing social welfare in the dynamic matching market. The key features of our model is a fixed rule-based function and a learnable component that can be trained by data gathered online with no prior knowledge or expert supervision. The output of the learnable component is a probability distribution over the matching decisions for the individual customers. 
The proposed hybrid model enables the learning algorithm to find an effective matching policy that simultaneously satisfies the customers' servicing preferences. 
The simulations show that the learning algorithm learns an effective matching policy for different generation-consumption profiles and exhibits better performance compared to standard online matching heuristics such as Match on Arrival, Match to the Highest, and Match to the Earliest Deadline policies.
\end{abstract}

\begin{IEEEkeywords}
Dynamic matching markets, power systems, reinforcement learning, policy gradient.
\end{IEEEkeywords}

\section{Introduction}  \label{sec:int}
\IEEEPARstart{L}{arge} scale integration of renewable energy sources (RES) into the power systems is a major strategy for the decarbonization of the energy system. 
RES include combinations of utility-scale centralized and distributed wind and solar power plants. 
Integration of RES in the operation and control of the grid is a significant challenge because photovoltaic (PV) solar and wind are highly uncertain, inherently variable, and largely uncontrollable. 
The flexibility of loads represent a promising solution to manage the uncertainty in RES and thereby achieve greater penetration of RES \cite{oikonomou2019deliverable}. 
The challenge here is the scheduling and matching of uncertain supply to the uncertain flexible load \cite{parag2016electricity}. 


\subsection{Background}

Traditionally, the power markets were managed using bulk load flexibility options such as load shifting and load curtailment, supplied by aggregators to the system operator, to manage the uncertainty of the real-time market \cite{parvania2014iso, khatami2018scheduling, muthirayan2020selling}. 
The drawback of the bulk load flexibility options is that (i) it limits the diversity of options that is available for the system operator by pooling the resources together in bulk and (ii) does not take into account the preferences of individual flexible loads.

Recently, many works have proposed peer-to-peer (P2P) energy trading markets. Peer-to-peer energy markets treat every local customer and local genenerator has an independent entity and thus can alleviate the shortcoming of the bulk load flexibility approach.
P2P energy trading markets can be classified under three categories: (i) centralized markets \cite{tushar2021peer, luth2018local}, (ii) decentralized markets \cite{morstyn2018bilateral, guerrero2018decentralized, khorasany2019decentralized, alashery2020blockchain, soriano2021peer}, and (iii) community markets \cite{paudel2018peer, moret2018energy, khorasany2021two}.
The P2P energy trading models proposed in \cite{luth2018local, morstyn2018bilateral, guerrero2018decentralized, khorasany2019decentralized, alashery2020blockchain, soriano2021peer, paudel2018peer, moret2018energy, khorasany2021two} focus on offline optimization-based solutions for trading in power system operation. 
Thus, these solutions can be very sub-optimal because they are not adaptable in real time operation of power markets. 
In addition to the solutions in \cite{luth2018local, morstyn2018bilateral, guerrero2018decentralized, khorasany2019decentralized, alashery2020blockchain, soriano2021peer, paudel2018peer, moret2018energy, khorasany2021two}, the works in \cite{wang2021surrogate, ye2021scalable, qiu2021scalable} propose learning-based solutions for online P2P energy trading.
In \cite{wang2021surrogate} the authors propose a simplified power market framework for P2P energy trading in multi-energy systems.
The primary focus of this study is on energy trading but not managing flexibility for RES integration, which is the focus of our study.
In \cite{ye2021scalable, qiu2021scalable} the authors propose a specific price based market framework for coordinating the prosumers in the market in order to minimize peak load. 
The limitation of these works is that the developed market framework is a coarse model that cannot be extended to other market scenarios.
In contrast, we propose an online solution for general market scenario with the objective to maximize social welfare. 

\subsection{Contribution}
This paper proposes a {\it dynamic matching market} for power market operation in local communities and a {\it Deep Reinforcement Learning (DRL)} solution to learn an online matching policy to match the random supply and flexible customers. 
A matching market is a P2P market that treats every participant, in this case customers with their load request and the local power supplies, as an independent entity. 
Thus, it can leverage customer preferences and availability to maximize the benefit to customers and the renewable integration. 
The market setting considered in this paper is an online power market, where the customers are flexible, their preferences are dynamic, and 
the future customer arrivals and renewable generation are variable and uncertain. 
The objective of the matching market is to maximize social welfare subject to servicing the customers using RES in such a setting. 
The proposed learning-based solution for such markets offers a flexible and efficient approach, which can learn without any prior experience or expert supervision or elaborate design. 

\begin{figure}[h]
\centering{
\includegraphics[width=0.8\linewidth]{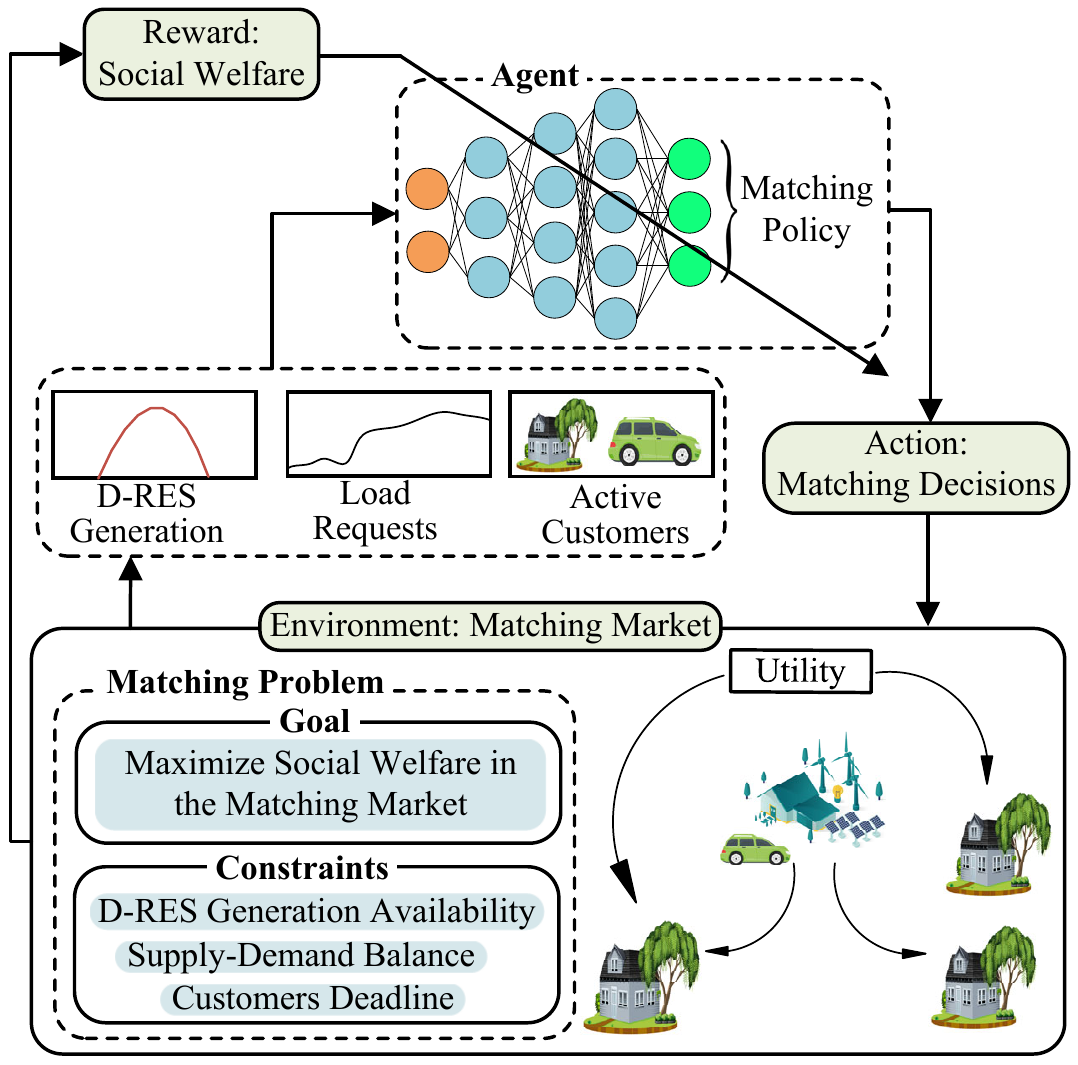}}
\caption{\small Schematic of the proposed learning-based solution for online matching in power system.} 
\label{Strucutre}
\end{figure}

The schematic of the proposed dynamic matching market model with the learning-based solution is shown in Fig. \ref{Strucutre}.
The proposed market model consists of a learning agent that outputs a matching policy for each time step.
The matching policy uses the information of local supply sources and customers (i.e., customers' load request and servicing constraints) to match the active customers with the available supply sources in the market.
After the market operation ends, the learning agent updates its matching policy based on the observed overall social welfare. 
Our detailed numerical studies show that unlike the existing online matching heuristics that deploy a fixed matching strategy in the market, the proposed learning solution is flexible to learn an effective matching strategy for a wide range of generation-consumption profiles, demonstrating its capability to match in real-time power system operation.

In summary, the key contribution of the paper is a learning solution for online matching in local communities that: a) can learn to match flexible customers with appropriate supply sources across load-generation scenarios in highly uncertain local online power markets, and b) requires no prior experience or expert supervision or elaborate design.

The rest of the paper is organized as follows: Matching market setting is described in Section \ref{sec:probform}. 
The proposed learning-based solution for online matching is presented in Section \ref{DRL}.
Parametric model for the matching policy is described in Section \ref{DRL}.
Numerical Studies are presented in Section \ref{Results} and the conclusions are drawn in Section \ref{Conls}.

\section{Matching Market Setting}
\label{sec:probform}

This section describes the matching market setting: the supply model, the customer model, the market state model, the dynamic matching problem and the performance metric.
 
\subsection{Energy Resources Model} \label{sec:supmod}
We consider two sources of supply for the dynamic matching market: 1) upstream grid supply (type gs), and 2) distributed renewable energy sources (D-RES) (type rs). 
We assume the market doesn't have a significant impact on the upstream grid.
Hence, the upstream grid supply, denoted by $p_t \in \mathbb{R}$, is sufficiently large and it is priced at the retail price of electricity, $c$/kWh. 
The D-RES, such as PV solar and wind generation, are by nature variable and uncertain, and their availability depends on weather, e.g., solar irradiance, wind speed. We denote the D-RES generated at time $t$ by $r_t \in \mathbb{R}$. The unit cost of renewable units is assumed to be zero. 
 
\subsection{Flexible Customer Model}
\label{sec:conmod}
\begin{figure}[h]
\centering{
\includegraphics[width=0.8\linewidth]{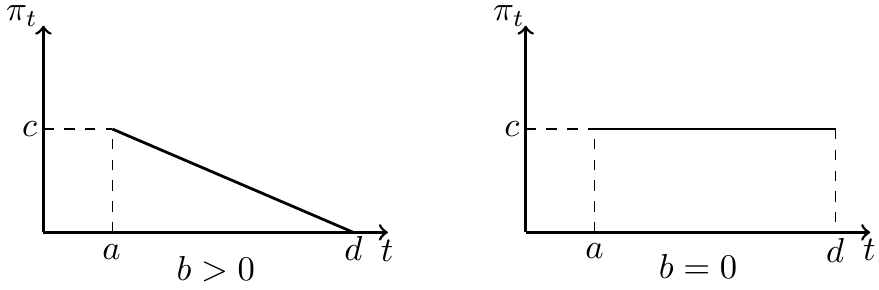}}
\caption{\small Illustration of utility function for different values of $b$. } 
\label{fig.Utility}
\end{figure}

Each customer (or load) is characterized by three parameters, $\{a^i,q^i,d^i\}$, where $a^i \in \mathbb{N}$ is the arrival time of the customer, $q^i \in \mathbb{R}$ is the load requested by the customer, and $d^i \in \mathbb{N}$ is the deadline by which the customer is to be served.
The heterogeneity of customers lies in the differing deadlines (also their criticality). 
Every customer arriving at the market (e.g., customer $i$) is characterized by its load request, arrival time $a^i$, servicing deadline $d^i$ and criticality rate $b^i$, at which its willingness to pay decreases from $a^i$ until $d^i$.
Hence, the utility function of customer $i$ representing its willingness to pay for a unit of energy can be defined as follows:
\begin{align} 
& \pi^i_t  = c - b^i(t-a^i), ~~~ \pi^i_t \geq 0, \ a^i \leq t \leq d^i, \nonumber \\
& b^i = \varphi c/(d^{i}-a^{i}),
\label{eq:ass-wtp}
\end{align}
where $\varphi \in [0,1]$ determines the reduction rate in customers willingness to pay for a unit of energy. 
The utility function for different values of $b$ is shown in Fig. \ref{fig.Utility}.

In Fig. \ref{fig.Utility}, customer's willingness to pay is less than or equal to the grid supply price $c$/kWh.
This is reasonable considering that the grid supply is available at this price at all times.
If $\varphi\!=\!1$, any remaining customer being served at the deadline will be willing to pay zero for the energy it receives.
On the other side, when $\varphi$ is set to 0, arriving customers can be served at any time by the market operator with no change to their willingness to pay.
The novelty of proposed customer model is its flexibility to capture a variety of servicing constraints for the active customers in the market, where customers' willingness to pay can remain fix or decay with time and at distinct rates.
We denote the number of customers who arrive on the platform at time $t$ by $n_t \in \mathbb{N}$, which is upper bounded by $\overline{n}$. 
We make this assumption because in real markets the number of customers will always be finite.


\subsection{Market State Model}\label{sec:scenmod}

Let $z_t := [a^\top_t, q^\top_t, d^{\top}_t, r_t]$\footnote{${[.]}^{\top}$ denote the matrix transpose operation.}, where $a_t \in \mathbb{N}^{\overline{n}}$ is the vector of the arrival times of the customers which arrive at time $t$, $q_t \in \mathbb{N}^{\overline{n}}$ is the vector of their respective requested loads, $d_t \in \mathbb{N}^{\overline{n}}$ is the vector of their respective deadlines, and $r_t \in \mathbb{R}$ is the amount of renewable generation at time $t$. The scenario at time $t$ is given by
\begin{equation}
Z^\top_t = [z^\top_1, z^\top_2, . . ., z^\top_{t-1}, z^\top_t]. \nonumber 
\end{equation}

The probability that $z_t = z$ is given by the stochastic process modeled by $\mathbb{P}\left(z_t = z \vert Z_{t-1}\right)$. This process is not known to the market operator. 
Let $x_t := [a^\top_t, q^\top_t, d^\top_t, q^\top_{p,t}, r_t]$, where $q_{p,t}$ denotes the vector of the portion of the requested loads that has not been served to the customers who arrived at $t$. We denote the set of all possible states at time $t$ by $\Omega_t$ and the state of the market by $X_t$. Then $X_t$ is given by
\begin{equation}
X^\top_t = [x^\top_1, x^\top_2, . . ., x^\top_{t-1}, x^\top_t]. \nonumber 
\end{equation}
Note that state $X_t$ depends on the scenario $Z_t$ and the matching decisions till time $t-1$.

\subsection{Dynamic Matching Problem}
The proposed dynamic matching market of the power system operates for a duration of $T$ within an epoch with time periods spaced equally at an interval $\Delta t$. The loads arrive in a sequential fashion and the generation of D-RES are governed by the stochastic process described in Section \ref{sec:scenmod}. At instant $t$, the market operator can decide to match the energy demands of the loads that are currently active for the increment of time $\Delta t$ to D-RES or the grid supply or wait till later to match it.

Denote the set of all the customers active on the platform at time $t$ by $A_t$. 
Denote the set of supply types by $S_t = \{gs, rs\}$. 
It follows that $A_t$ is a function of $X_t$. 
Let $q^i_p$ denote the portion of the requested load by customer $i$ that has not been served. Denote the amount of supply of type $j$ matched to customer $i$ at time $t$ by $M_{t}(j,i,X_t) \in \mathbb{R}$. 

The market operator has the following information to make its matching decisions at time $t$: the set of active customers $A_t$, their arrival time and deadlines, the amount of renewable generation $r_t$, customers active before $t$ and renewable generation for each time step before $t$. Let $c_j$ denote the unit cost of type $j$ supply. 
The dynamic matching problem \eqref{eq:matchingproblem} for the market operator is maximizing social welfare under the above information setting subject to the load and supply constraints:
\begin{align}
& \text{sup}_{M_{t}} \sum_{t = 1}^T \sum_{i \in A_t} \sum_{j \in S_t} (\pi^i_t - c_j)M_{t}(j,i,X_t), \ \text{s.t.} \nonumber \\
& \sum_{j \in S_t} M_{t}(j,i,X_t) \leq q^i_p, \ \forall i \in A_t, \ \sum_{i \in A_t} M_{t}(\text{rs},i,X_t) \leq r_t, 
\label{eq:matchingproblem}
\end{align}
where the dependency of $M_t$ on $X_t$ accounts for the dependency of the matching decision on the full state information. 
The problem as stated is an online problem because at any point of time $t$ during the operation of the market the market operator can only make its matching decision based on the information of the current and the past customers and supply quantities and with uncertainty on the future.
Here we have not included networks constraints assuming that they are not violated \cite{hayes2020co}.

\subsection{Performance Metric for a Matching Policy}

The performance measure for a general online matching policy, $\chi$ (whose policy for time $t$ is denoted by $\chi_t(.,.,.)$), for the dynamic matching problem is given by
\begin{align}
& W[\chi] = \sum_{t = 1}^T \sum_{i \in A_t} \sum_{j \in S_t} (\pi^i_t - c_j)M_{t}(j,i,X_t), \nonumber\\
& M_{t}(j,i,X_t) = \chi_k(j,i,X_t), \nonumber
\end{align}
which equals the social welfare achieved by the algorithm $M$ for the duration $T$ of the market. 

\section{Reinforcement Learning Model}\label{DRL}

Unlike the matching heuristics proposed in \cite{mehta2005adwords, jaillet2013online, blum2006online, muthirayan2020online}, we propose a learning-based solution that consists of (i) a trainable matching policy and (ii) a reinforcement learning algorithm to train the policy from the load and generation data for multiple instances (multiple epochs) of the market. 
The proposed reinforcement learning algorithm is a policy gradient algorithm that trains the matching policy from customer and D-RES data by evaluating its own performance. 
Thus, the proposed training procedure does not require supervision or expert knowledge. 
Next, we first briefly discuss the structure of the matching policy and then discuss the learning algorithm. We denote the expectation with respect to all sources of randomness by $\mathbb{E}[.]$.

\subsection{Matching Policy Preliminaries}
We denote the online matching policy by $\chi = \{\chi_1, \chi_2, \chi_3,...,\chi_T\}$, where $\chi_t$ is the discrete matching policy for time $t$. We define a discrete matching policy to be the policy that indicates whether a customer is to be matched to a supply or not without specifying the amount of matching. 
Let us denote the space of discrete matching at time $t$ by $\mathcal{M}_t$, where each member of this set is a feasible discrete matching for a state that can be realized at time $t$. 
For any $m \in \mathcal{M}_t$, the component that specifies whether a customer $i$ at $t$ is matched to supply type $j$, $m_{j,i} \in \{0,1\}$, where one denotes ``matched" and zero denotes ``not matched". Then a general matching policy $\chi_t$ is given by:
\begin{equation}
\chi_t : \Omega_t \rightarrow \mathcal{M}_t. \nonumber
\end{equation}

Aside the fact that the matching problem is an online decision problem with the future ridden by uncertainties, the problem has several challenges from a reinforcement learning point of view. The challenges we are about to list are common to reinforcement learning in general \cite{dulac2021challenges}. Firstly, {\it the action space of the matching decision is large and specifically exponential in the number of customers}. For example, if there are $m$ supplies and $n$ customers, then there are $m^n$ ways of matching; thus it is exponential in the number of active customers. Secondly, not all actions from this space are feasible. For example, if one supply is limited then it cannot be matched to all the active consumers. As a second example, customer's servicing constraint such as deadline restricts the action space. Thus, some {\it actions are infeasible and their infeasibility is state dependent}. Thirdly, {\it reinforcement learning can converge to a local optimum}, which is a general challenge and so applies to the matching problem as well. All of this combined rises the following question, {\it what is a suitable matching model that is simple and efficient to learn, simultaneously satisfies the action constraints and still has the capacity for RL to converge to a good solution.} We propose a reinforcement learning model to address this problem.

We do not overview the RL setting and refer the reader to review papers on DRL for the formal description of the RL setting and the literature on DRL \cite{sutton2018reinforcement, arulkumaran2017deep, zhang2018review, nguyen2020deep}. Our model is based on the following intuition. We simplify the output of the policy to be trained by RL to just ``match" or ``not to match" for each active customer, ignoring which supply to match and whether the action is feasible or not. Thus, the action space of the output of the component that is trained is linear in the number of active consumers. Given this output, the final matching to the supplies still has to be decided. While the above approximation simplifies the action space, depending on how the final matching is decided, the final matching can be grossly simplified and very restrictive. As an example, ``to match" can be interpreted as match to the grid always. Since the matching is always restricted to the grid, the algorithm will be forced to learn only within the scope of matching to the grid, severely restricting the final outcome. Thus, while the simplification reduces the complexity of the model, it can severely handicap the learning and worsen the outcomes. 

The proposed matching policy in this study outputs a discrete matching decision.  
Once the discrete matching is computed, market operator operationalizes the discrete matching by computing the actual power to be matched, i.e., $M_t$. The splitting of the decision as a discrete matching and then the computation of the matching $M_t$ simplifies the decision space across which the learning algorithm will have to explore, since, with this simplification, the learning algorithm will have to explore only to ``match" or ``not to match". Moreover, this also reduces the complexity of the model by significantly reducing the dimension of the decision from exponential in the number of active customers to linear.

\subsection{Model of Matching Policy}

We propose a certain structure for the matching policy $\chi_t$, which has a learnable component and a fixed component. The fixed component accounts for the fact that the customers have to be matched before their deadline and that the matching to renewable generation cannot exceed the amount that is generated. 
The learnable component is the policy $\mu_t$. The policy $\mu_t$ outputs the probability of matching a load without specifying which type of supply to match to. This simplifies the learnable part of the matching policy, whose size would otherwise scale exponentially with the number of customers.
Denote the space of such matching decisions for time $t$ by $\mathcal{H}_t$. 
For any $m \in \mathcal{H}_t$, the component that corresponds to customer $i$, $m_i \in \{0,1\}$, where zero denotes ``not to be matched" and one denotes ``to be matched". 
We denote the space of probability measures over the set $\mathcal{H}_t$ by $\mathcal{P}_{\mathcal{H}_t}$. Then, the policy $\mu_t$ is given by:
\begin{equation}
\mu_t : \Omega_t \rightarrow \mathcal{P}_{\mathcal{H}_t}. \nonumber
\end{equation}

Let $m_t \in \mathcal{H}_t$ be given by $m_{t} \sim \mu_t$. We denote the component of $m_t$ that corresponds to customer $i$ by $m_{t,i}$, where $m_{t,i} \in \{0, 1\}$. The output $m_k$ is input to a second function, $\varphi$, which matches the customers for whom $m_{t,i} = 1$ and up to $r_t$ of them to the renewable units. It matches any remaining customer for whom $m_{t,i} = 1$ to the grid supply. If $\sum_i q_i m_{t,i} < r_t$ then in addition to the above step it matches the remaining $r_t - \sum_i q_i m_{t,i}$ to available customers. This function is the key function that influences the learning algorithm to converge to an effective solution. 
The function $\varphi$ is given by:
\begin{equation}
\varphi : \mathbb{R} \times \mathcal{H}_t \rightarrow \mathcal{M}_t. \nonumber
\end{equation}

Denote the component of $\varphi$ that specifies whether customer $i$ is matched to supply type $j$ by $\varphi_{j,i}$. The output of $\varphi$ is input to a third function, $\nu$, that overturns the decision to not to match the customers with immediate deadline:
\begin{equation}
\nu_{j,i} = \left\{ \begin{array}{cc} 1 & \text{if} \ d^i = t, \ \text{$i$ is active}, \ \varphi_{rs,i} = \varphi_{gs,i}= 0 \\ & \text{and} \ j = \text{gs}, \\
\varphi_{j,i} & \ \text{otherwise}. \end{array} \right. \nonumber
\end{equation}
The function $\nu$ ensures that the customers are served by their deadline. 
Thus, the overall matching policy for time $t$, $\chi_t$, is given by:
\begin{equation}
\chi_t = \nu \circ \varphi \circ m_t, m_t \sim \mu_t.
\label{eq:matchingalg}
\end{equation}
Let $\mu_t$ to be parameterized by $\theta_t$ and denote the parameterization by $\mu_t(.;\theta_t)$. 
Later we discuss the parametric model for $\mu$, $\mu = \{\mu_1, \mu_2, \mu_3,..., \mu_T\}$. The learning algorithm trains the parameter $\theta_t$ for every $t$ using the load and generation data by evaluating its own performance. 

Let $X = \{X_1, X_2, X_3,..., X_T\}$. We use $X \sim \mu$ as shorthand notation to denote the dependence of the state on the random policy $\mu$. Given the performance metric for the matching policy, the value function for the policy $\chi$, $V_\chi$, is the expected welfare for the duration $T$ of the market. Thus:
\begin{align}
& V_\chi = \mathbb{E}_{X\sim\mu} [W[\chi]] \ \text{where} \nonumber \\
& W[\chi] = \sum_{t=0}^T \sum_j \sum_{i\in A_t} (\pi^i_t - c_j) \chi_{t,j,i}(X_t).
\end{align}

\subsection{Policy Gradient Learning Algorithm}

\subsubsection{Gradient Estimate}

We first derive a gradient estimate for the matching policy that is calculable from the data observed over the period of the market. 

We use $\mathbb{E}_{X_t \sim \mathbb{P}_t(.)}$ as a shorthand for expectation over $X_t \sim \mathbb{P}(. \vert X_{t-1}, \chi_{t-1})$, where $\mathbb{P}(. \vert X_{t-1}, \chi_{t-1})$ denotes the transition probability from state $X_{t-1}$ under the policy $\chi_{t-1}$. Let $m_{t:T} = \{m_t, m_{l+1},...,m_T\}, \mu_{l:T} = \{\mu_t, \mu_{l+1},..., \mu_T\}$. Let
\begin{align}
& V^\chi_{t+}(X_{t+1}) := \mathbb{E}_{m \sim \mu} \sum_{l=t+1}^T \left[\sum_j \left[(\pi_l(A_l) - c_j)^\top \chi_{l,j,A_l}\right]\right], \nonumber \\
& = \mathbb{E}_{m_{t+1:T}\sim \mu_{t+1:T}}\sum_{l=t+1}^T \left[\sum_j \left[(\pi_l(A_l) - c_j)^\top \chi_{l,j,A_l}\right]\right], \nonumber
\end{align}
where $(\pi_l(A_l) - c_j)^T \chi_{l,j,A_l}$ is the shorthand for $\sum_{i\in A_t} (\pi^i_t - c_j) \chi_{t,j,i}(X_t)$. Let
\begin{align}
& v^\chi_t := \sum_j\left[ (\pi_k(A_t) - c_j)^\top\chi_{k,j,A_t}(X_t)\right], \nonumber \\
& V^\chi_t(X_t):= \mathbb{E}_{m_t\sim \mu_t}\left[V^\chi_t(X_t\vert m_t)\right], 
\end{align}
where, $V^\chi_t(X_t\vert m_t) := v^\chi_t + \mathbb{E}_{X_{t+1} \sim \mathbb{P}_{t+1}(.)}\left[V^\chi_{t+}(X_{t+1})\right]$.
Then, from the definition of $V_\chi$ it follows that
\begin{equation}
\frac{\partial V_\chi}{\partial\theta_t} = \mathbb{E}_{X_t}\left[\frac{\partial V^\chi_t(X_t)}{\partial\theta_t}\right]. \nonumber
\end{equation}
Then, from the definition of $V^\chi_t(X_t)$ it follows that
\begin{align}
& \frac{\partial V_\chi}{\partial\theta_t} = \mathbb{E}_{X_t}\sum_{m_t \in \mathcal{H}_t}\frac{\partial\mu_{t}(m_t;\theta_t)}{\partial \theta_t} \left[v^\chi_t \right. \nonumber\\
& \left. + \mathbb{E}_{X_{t+1} \sim \mathbb{P}_{t+1}(.)} V^\chi_{t+}(X_{t+1})\right]. \nonumber
\end{align}
We can rewrite the above equation as
\begin{equation}
\frac{\partial V_\chi}{\partial\theta_t} = \mathbb{E}_{X_t, m_t \sim \mu_t} \left[\frac{\partial \log\mu_{t}(m_t;\theta_t)}{\partial \theta_t} V^\chi_t(X_t\vert m_t)\right]. 
\label{eq:grad-V-theta-k}
\end{equation}

The previous equation gives the gradient of the value function w.r.t the policy parameter $\theta_t$. 
An unbiased estimate of the gradient in \eqref{eq:grad-V-theta-k} is:
\begin{equation}
    \delta^\theta_t = \left[\frac{\partial \log\mu_{t}(m_t;\theta_t)}{\partial \theta_t} V^\chi_t(X_t\vert m_t)\right].
    \label{eq:stoch-grad-critic}
\end{equation}
The gradient in \eqref{eq:stoch-grad-critic} is not computable because $V^\chi_t(X_t\vert m_t)$ is unknown. In place of $V^\chi_t(X_t\vert m_t)$, we can use the social welfare from $t$ to $T$ for a sample epoch under the policy $\chi$:
\begin{equation}
\delta^\theta_{t,r} = \frac{\partial \log\mu_{t}(m_t;\theta_t)}{\partial \theta_t} \left[\sum_{l=t}^T v^\chi_l \right].
\label{eq:stoch-grad-reinforce}
\end{equation}
This gradient is computable using the data from a sample epoch $\left(Z = \{Z_1,Z_2,...,Z_T\}\right)$ and the matching decisions as determined by the policy $\chi$ for this sample epoch. Additionally the gradient estimate is also unbiased because
\begin{equation}
\frac{\partial V_\chi}{\partial\theta_t} = \mathbb{E}[\delta^\theta_{t,r}]. \nonumber
\end{equation}

\subsubsection{Vanilla Policy Gradient Learning Algorithm}
The vanilla policy gradient learning algorithm learns the policy parameter for each time step by the following stochastic gradient ascent algorithm:
\begin{equation}
\theta_t \leftarrow \theta_t + \gamma_\theta \delta^\theta_{t,r}, 
\label{eq:grad-update}
\end{equation}
that iteratively updates $\theta_t$ by using the computed gradient $\delta^\theta_{t,r}$ for multiple sample epochs, one every update step. 

\subsubsection{Actor-Critic Policy Gradient Learning Algorithm}

We also propose a second learning algorithm, called the actor-critic algorithm AC$-k$. This algorithm, in addition to learning the matching policy $\mu$ also learns an approximation of the value function $V^\chi_t(X_t)$ called the critic function. The critic function is parameterized by $\phi_k$ and is denoted by $V^\chi_k(X_k;\phi_k)$. The approximate policy gradient for the actor-critic algorithm AC$-k$ is given by:
\begin{equation}
\!\!\delta^\theta_{t,k} = \!\frac{\partial \log\mu_{t}(m_t;\theta_t)}{\partial \theta_t} \!\!\left[\sum_{l=t}^{t+k-1} v^\chi_l + V^\chi_{t+k}(X_{t+k};\phi_{t+k})\!\right]\!.  
\label{eq:stoch-grad-actor-critic}
\end{equation}
The actor-critic algorithm (AC$-k$) learns the policy parameters by the following stochastic gradient ascent algorithm:
\begin{equation}
\theta_t \leftarrow \theta_t + \gamma_\theta \delta^\theta_{t,k}. 
\label{eq:grad-update-actor-critic}
\end{equation}
The parameter $\phi_k$ of the critic function is similarly learnt by stochastic gradient descent for its least-squares error:
\begin{equation}
\phi_k \leftarrow \phi_k - \gamma_\phi \left(V^\chi_{k}(.;\phi_k) - \sum_{l=k}^T v^\chi_l\right).
\label{eq:critic-update}
\end{equation}
The complete algorithm is given in Algorithm \ref{alg:policygradient}. The algorithm uses the ADAM gradient algorithm \cite{kingma2014adam} of the gradient updates in \eqref{eq:grad-update-actor-critic} and \eqref{eq:critic-update}. 

\begin{algorithm}[]
\begin{algorithmic}[1]
\STATE \textbf{Initialize}: $\mathcal{D} = \varnothing$, $j = 0$
\STATE \textbf{Initialize}: $\theta_k ~ \forall ~ k \in [1,\ldots,T]$. $N$: number of epochs
\FOR {$i = 1,\ldots,N$}
\STATE Increment $j = j + 1$
\STATE Set $D_i = ((X_k,m_k,v^\chi_k) ~ \forall ~ k \in [1,\ldots,T])$
\STATE Include $D_i$ into $\mathcal{D}$
\IF{j == M}
\STATE Update $\theta_k$ by ADAM of Eq. \eqref{eq:grad-update-actor-critic} using $\mathcal{D}$
\STATE Update $\phi_k$ by ADAM of Eq. \eqref{eq:critic-update} using $\mathcal{D}$
\STATE j = 0; $\mathcal{D} = \varnothing$
\ENDIF
\ENDFOR
\end{algorithmic}
\caption{\bf Actor-Critic (AC$-k$) Policy Gradient Learning Algorithm}
\label{alg:policygradient}
\end{algorithm}

\section{Parametric Model for the Matching Policy: Temporal Convolution Neural Network}
\begin{figure}[h]
\centering{
\includegraphics[width=1\linewidth]{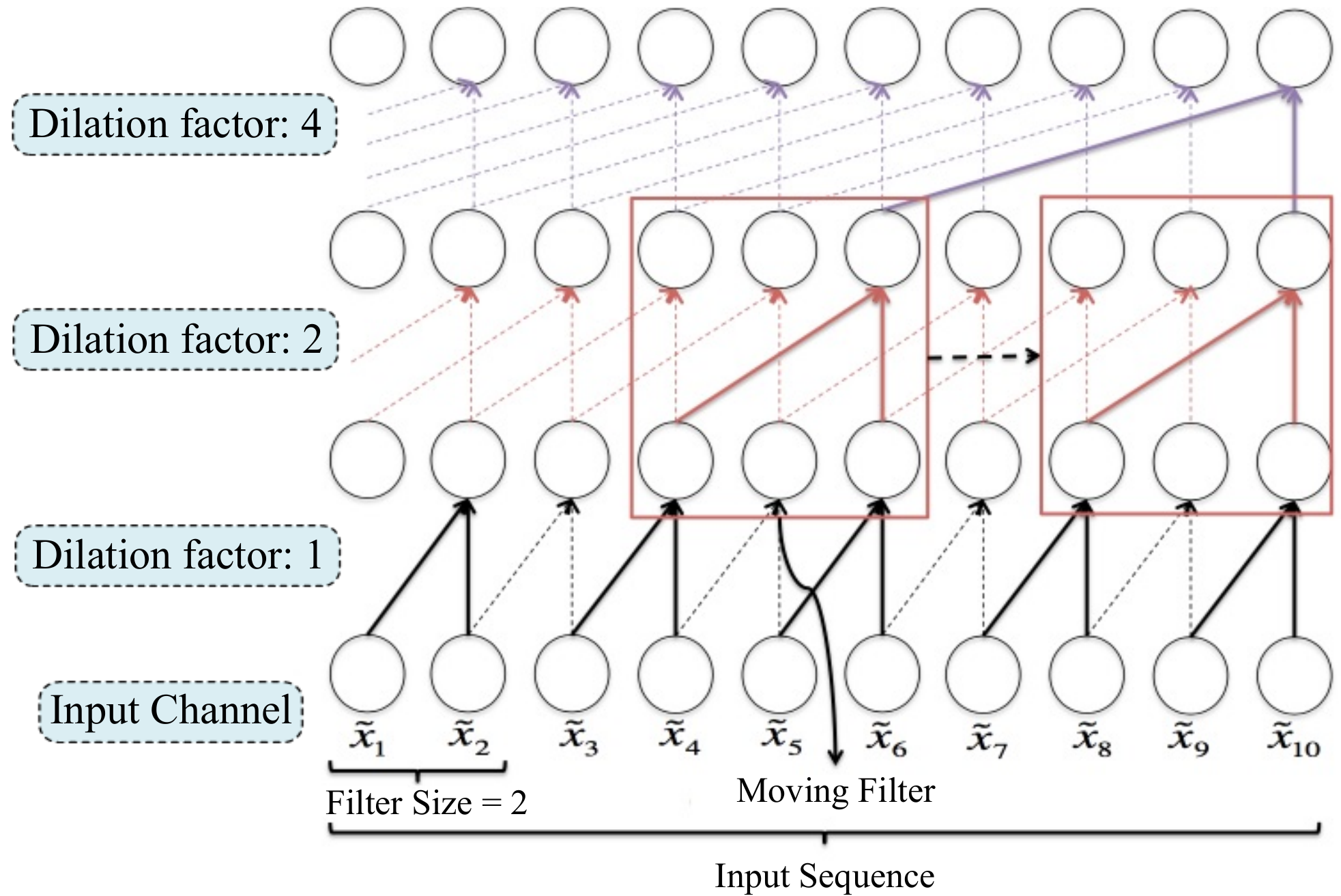}}
\caption{\small Temporal convolution network.} 
\label{fig.TCN_New}
\end{figure}

The proposed parametric model for the learnable component $\mu$ is a Temporal Convolution Network (TCN) \cite{bai2018empirical}, whose weights are collectively denoted by $\theta$. 
TCNs use $1$D fully-causal convolutional network (FCN) architecture (see Fig. \ref{fig.TCN_New}). TCNs are specified by (i) the number of input channels, (ii) the number of layers or blocks, (iii) the number of filters (similar to the number of filters in Convolutional Neural Networks (CNNs)), (iv) the filter size, and (v) the dilation factor for each layer or block. In Fig. \ref{fig.TCN_New}, the number of input channels is one, the number of filters or convolution operations is one for each layer (where the different filters are color coded differently), the number of inputs for each filter operation or the filter size is two, the dilation factor varies by layers and the dilation factor is doubled in each subsequent layer. 

\begin{figure}[h]
\centering{
\includegraphics[width=0.5\linewidth]{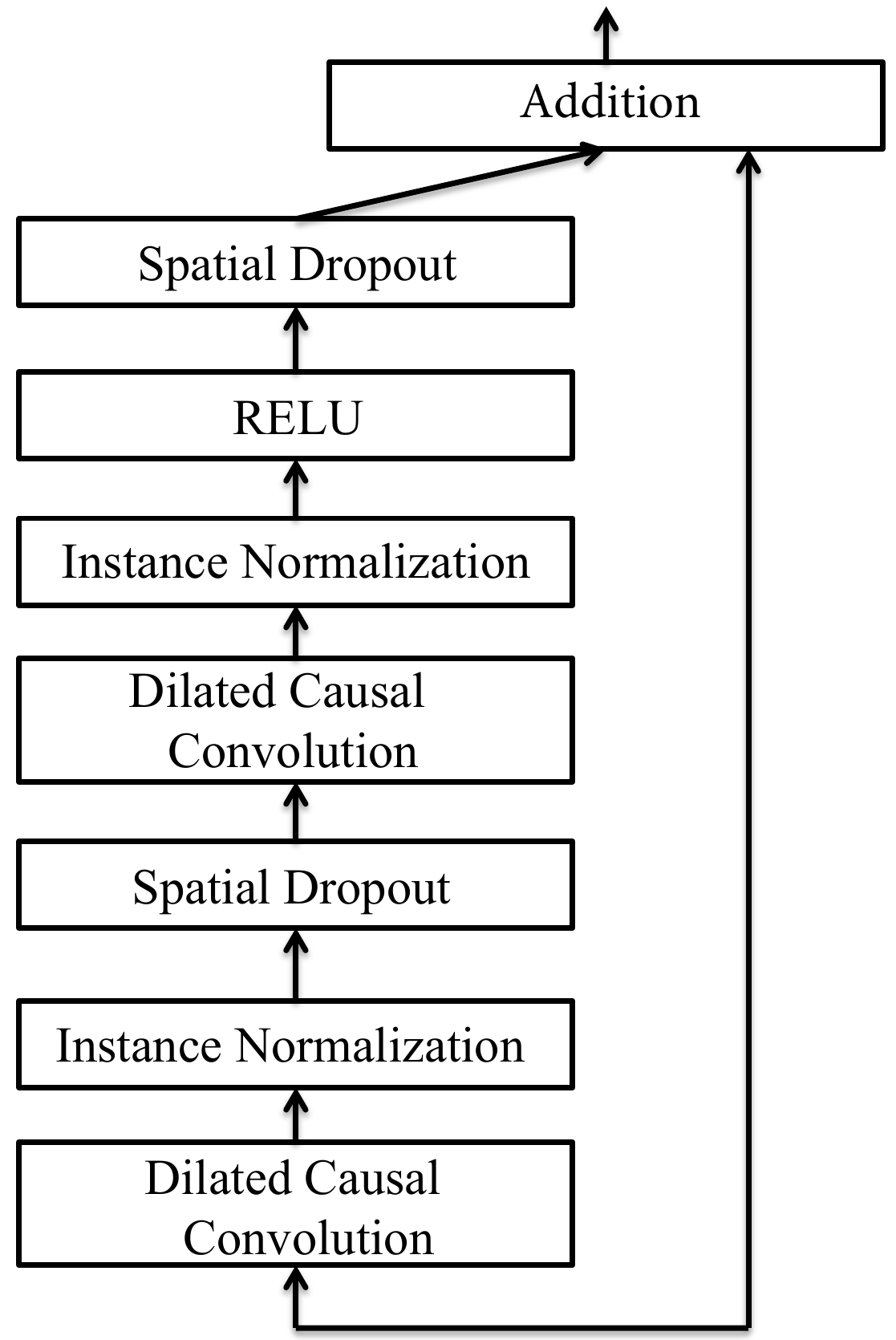}}
\caption{\small A single block of TCN.} 
\label{fig.TCN-Block_New}
\end{figure}

A simple causal convolution can only look back at history of size that is linear with the depth of the network. This makes it challenging to apply simple causal convolutions on sequence tasks. The use of dilation in the upper layers allows the architecture to look back at history of size that is exponential in the depth of the network, making the model efficient. The inclusion of residual connections enable the construction of deep networks as in CNNs. 
Each filter of a TCN corresponds to the kernel of the convolution operation it corresponds to and each convolution operation is 1D (convolution over the time dimension). Within each layer or block there can be multiple sub-layers of causal convolution layers (see Fig. \ref{fig.TCN-Block_New}). Instance Normalization \cite{ulyanov2016instance} and Spatial Dropout \cite{srivastava2014dropout} can also be included in each layer (or block). Normalization is a standard technique to improve the speed of convergence and dropout is a widely used technique to regularize models in deep learning. 

Just as in a recurrent neural network (RNN), the output of a hidden layer is computed by applying the filters of a layer repeatedly by shifting them by a stride of one over the sequence produced by the previous layer (see Fig. \ref{fig.TCN_New}). Causality is achieved by ensuring the inputs to the filter from the input sequence are limited to the step in the output sequence of the layer for which the output is being calculated (see Fig. \ref{fig.TCN_New}).


Temporal convolution networks have several advantages. Firstly, TCNs can model input lengths of variable size just like RNNs. 
Unlike RNNs, TCNs do not have the problem of exploding or vanishing gradients because the backpropagation path in TCNs is different from the temporal direction. 
TCNs, by their very construction, are very flexible models because they provide multiple ways to expand the receptive field, for eg., either by using larger dilation, or adding additional layers, or just by increasing the size of the kernel filters. Additionally, computation is efficient in TCNs because convolutions in an individual layer are parallelizable. Thus, TCNs combine the best aspects of both RNNs and CNNs and are highly effective at modeling sequential data.

\subsection{TCN Model for $\mu$}

Let the TCN corresponding to $\mu$ be functionally denoted by TCN$_\mu$. Let $\tilde{x}_t = [q^\top_t, d^\top_t, q^\top_{p,t}, r_t]$. Let $\tilde{X}^\top_t = [\tilde{x}^\top_1, \tilde{x}^\top_2, \tilde{x}^\top_3,...,\tilde{x}^\top_t]$. For each time step $t$, the input to the TCN is $\tilde{X}_t$ and the output of TCN is the individual probabilities that the respective customers will be matched. Let the vector of probabilities that the customers will be matched be given by $P^m_\mu \in [0,1]^{\overline{n}\times T}$. Then:
\begin{equation}
P^m_\mu = \text{TCN}_\mu(\tilde{X}_t). \nonumber
\end{equation}

The input $\tilde{X}_t$ is fed to the TCN model as shown in Fig. \ref{fig.TCN_New}, i.e. the input $\tilde{X}_t$ is fed to the NN as the sequence $\{\tilde{x}_1, \tilde{x}_2, \tilde{x}_3,...,\tilde{x}_t\}$. Clearly for this setting, the number of input channels is equal to the size of $\tilde{x}_k$ for any $k$, which is equal to $\overline{n}\times 3+1$. 
The exact specification of the filter size, the number of filters, the number of layers or blocks are discussed later in the results section. The dilation factor is increased by two for every subsequent layer. 

The output of the TCN model is of a fixed size for each time step $t$. The size is the maximum number of customers that can be active at any point of time, which is equal to $\overline{n}\times T$. While reading the output of the TCN model at time step $t$, the values of probability of matching corresponding to the active customers are read. Denote the probability of matching for the active customer $i$ at time step $t$ by $P^m_{\mu,i}$. The distribution $\mu_t$ is constructed using these values:
\begin{equation}
\mathbb{P}(m_{t,i} = 1) = P^m_{\mu,i}, \ \mathbb{P}(m_{t,i} = 0) = 1 - P^m_{\mu,i}. \nonumber
\end{equation}

\subsection{Policy Gradient Algorithm for the TCN Model}

For the TCN model, the weights are given by the weights of the kernel filters. Since the same kernel filters are applied at each time step, $\theta_t = \theta$, where $\theta$ denotes the weights of all the kernel filters. Thus, the vanilla policy gradient algorithm for this model is given by:
\begin{equation}
\theta \leftarrow \theta + \gamma_\theta \sum_t \delta^\theta_{t,r} \nonumber.
\end{equation}
Similarly the actor-critic algorithm AC$-k$ is given by:
\begin{equation}
\theta \leftarrow \theta + \gamma_\theta \sum_t \delta^\theta_{t,k} \nonumber.
\end{equation}

\section{Results and Discussion}\label{Results}

\begin{figure}[h]
\centering{
\includegraphics[width=1\linewidth]{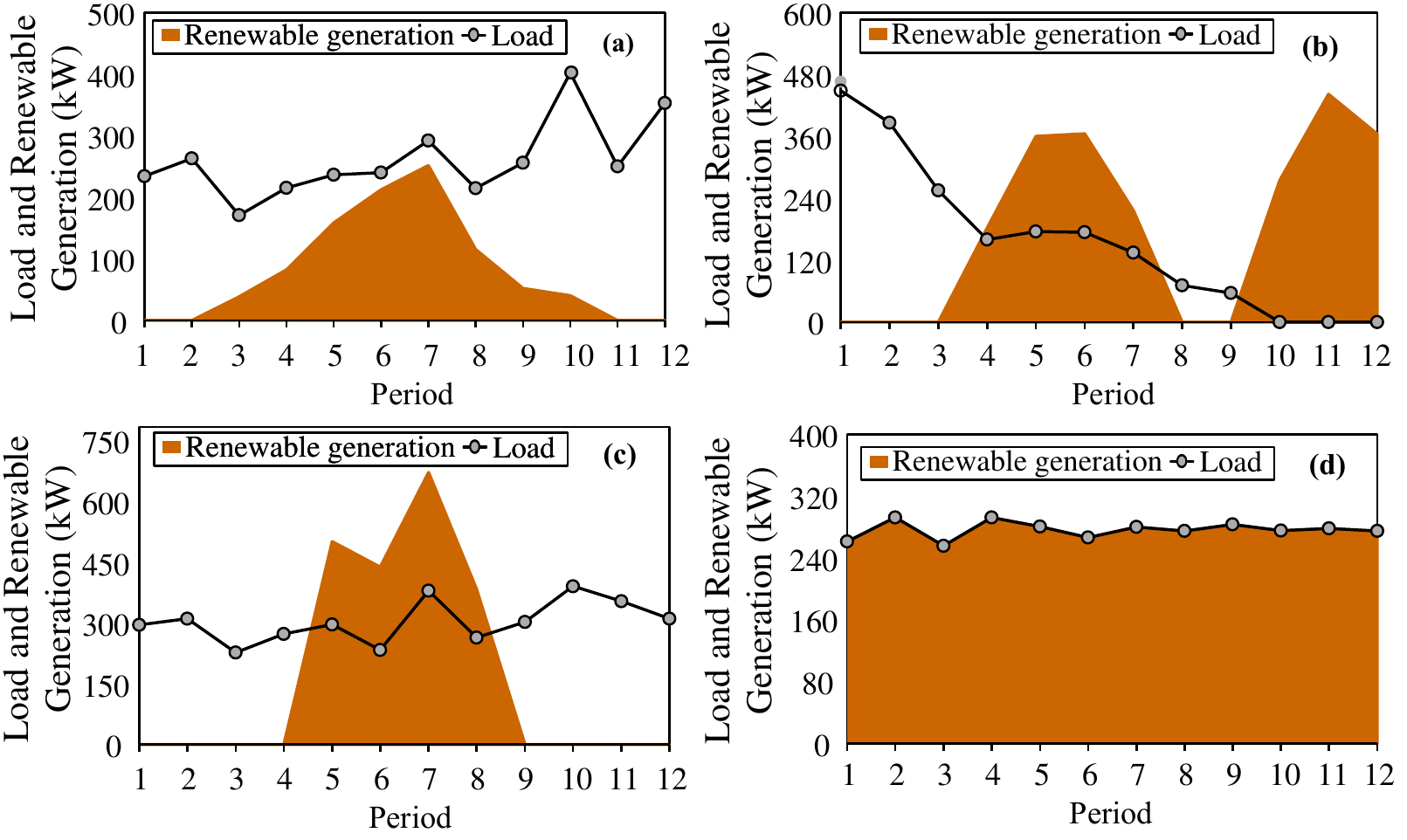}}
\caption{\small Load request and renewable generation: (a) Scenario 1, (b) Scenario 2, (c) Scenario 3, (d) Scenario 4.} 
\label{fig.Load_Profile}
\end{figure}

In this section we analyse the performance of the learning algorithm across different generation-consumption profiles. We also discuss specific market outcomes for each scenario to demonstrate how the learning algorithm matches. 
To evaluate the performance of the learning algorithm, the following online matching algorithms are considered:
\begin{itemize}
\item Match on Arrival (MA) algorithm: this algorithm matches the available renewable sources to the arriving customers. In case the renewable source is insufficient, the remaining customers would be matched to the grid supply.  
\item Match to the Highest (MH) algorithm: this algorithm matches the available renewable supply to the customers with the maximum willingness to pay value. Any remaining customer with an immediate deadline would be matched to the grid supply. 
\item Match to the Earliest Deadline (MED) algorithm: this algorithm matches the available renewable supply to the customers with the earliest deadlines. If the renewable supply is not sufficient enough, the customer with an immediate deadline would be matched to the grid supply. 
\item Learning Algorithm 1 (LA1): it denotes the vanilla policy gradient learning algorithm.
\item Learning Algorithm 2 (LA2): it denotes the actor-critic algorithm of type AC-1.
\item Offline Optimal Algorithm (OOA): this is the optimal matching solution calculated in hindsight for the actual realization of the load requests and renewable generation over the period of the market.
\end{itemize}
The performance of the developed algorithms is compared in terms of the average social welfare across a certain number of epochs. 
The best hyper-parameters for TCN and Adam optimizer were selected by choosing the best parameters' combination across the scenario.
The best hyper-parameters for the TCN model were identified to be 3 for the number of blocks, 4 for the number of filters, 3 for the filter size, 0.1 for the dropout factor and 4 for the dilation factor. 
Sigmoid function was utilized as the activation function for each output of TCN.
We used the following values for the parameters of the ADAM algorithm: ${\alpha}=0.75$, ${\beta}_1=0.9$, ${\beta}_2=0.999$, ${\epsilon}=10^{-8}$, where $\alpha$ is the leaning rate and ${\beta}_1$, ${\beta}_2$ are exponential decay rates for the moment estimates. 
The best batch size is 80 and 120 for LA1 and 20 for LA2.
The simulations are carried out on a sample power system with average load and renewable generation data for five different loading and generation scenarios as follows:
\begin{itemize}
    \item Scenario 1: This scenario considers customers characterized by random short deadlines, with an average deadline of 4 time periods from arrival, and limited generation during the middle of the epoch. The average load request and renewable generation for this scenario are shown in Fig. \ref{fig.Load_Profile}-(a).  
    \item Scenario 2: In this scenario, customers who arrive earlier have larger deadlines, while others who arrive during the middle of the epoch have shorter deadlines. The average load request and renewable generation for this scenario are shown in Fig. \ref{fig.Load_Profile}-(b).
    \item Scenario 3: This scenario considers customers characterized by random large deadlines, with an average deadline of 8 time periods from arrival and excess generation during the middle of each epoch. The average load request and renewable generation for this scenario are shown in Fig. \ref{fig.Load_Profile}-(c). 
    \item Scenario 4: This scenario considers customers characterized by fixed large deadlines of 8 time periods from arrival. The average load request and renewable generation for this scenario are shown in  Fig. \ref{fig.Load_Profile}-(d).
    \item Scenario 5: This scenario is a hybrid of scenarios $1, 2$ and $3$, where on every $3k+1$th epoch, $3k+2$th epoch, $3(k+1)$th epoch, $k \in \{0,1,2,...\}$, the load and generation and customer characteristics correspond to scenarios $1, 2$ and $3$, respectively.
\end{itemize}

In the first four scenarios, LA2 is fed with load and generation data for 200 epochs, while LA1 and rest of the algorithms are fed with load and generation data for 800 epochs, generated using normal distribution with the average values shown in Fig. \ref{fig.Load_Profile}.
The standard deviation to produce load and renewable generation profiles for scenarios 1-3 is equal to 15\% of the average load and generation, shown in Fig. \ref{fig.Load_Profile}, while it is 50\% of the average load and generation for scenario 4 to consider severe uncertainty in local load and generation.
Every epoch includes 12-period load and generation samples and considering the number of epochs discussed above, $(200~\text{epochs}) \!\times \! (12~\text{periods}) = 2400$ load and generation samples are utilized for LA2, while $(800~\text{epochs}) \!\times \! (12~\text{periods}) = 9600$ samples are utilized for LA1 and other algorithms in every scenario.


\subsection{Numerical Results}

The arriving load request and renewable generation are random in all the scenarios, but their mean level varies with time and differently in each scenario, as shown in Fig. \ref{fig.Load_Profile}.
The average social welfare that is achieved by each algorithm for each scenario is shown in Table \ref{tab:Numerical_results}. 
The results presented for the learning algorithms for each scenario are the best of the scores across the multiple runs. 
We find that the learning algorithms of the both types achieve the highest average social welfare among the online solutions across most of the scenarios. 
While the algorithms MH, MA and MED achieve an average score of $143.4, 151.8$ and $132.4$, the learning algorithm LA1 achieve an overall average of $\bf{170.8}$ and the learning algorithm LA2 achieve an overall average of $\bf{173.1}$, which is closest to the OOA. 
We also note that LA1 and LA2 are the top performing online algorithms for all the scenarios except scenario $1$.
While LA1 and LA2 achieve an almost similar score across the scenarios, the key difference is that the algorithm LA1 requires 800 epochs of data, while the algorithm LA2 requires just $200$ epochs of data to achieve the same level of performance. 

\begin{table}[h]
\caption{\small Average Social Welfare}
\label{tab:Numerical_results}
\centering
\resizebox{\columnwidth}{!}{\begin{tabular}{ccccccc}
\hline
\textbf{Scenario\textbackslash{}Algorithm} & \textbf{MA} & \textbf{MH} & \textbf{MED} & \textbf{LA1} & \textbf{LA2} & \textbf{OOA} \\ \hline
\textbf{Scenario 1 (\$)} & {\bf 150.4} & 43 & 28.4 & {\bf 130.5} & 126.4 & 150.4 \\
\textbf{Scenario 2 (\$)} & 84.7 & 136.7 & 118.6 & {\bf 137.4} & 134.7 & 149.6 \\
\textbf{Scenario 3 (\$)} & 160 & 178.7 & 174.3 & 188.5 & {\bf 202.1} & 256 \\
\textbf{Scenario 4 (\$)} & 232.7 & 262.2 & 260.2 & 266.9 & {\bf 267.1} & 315.3 \\
\textbf{Scenario 5 (\$)} & 131.6& 96.6 & 80.5 & 130.8 & {\bf 135.3} & 162 \\
\textbf{Average (all, \$)} & 151.8 & 143.4 & 132.4 & 170.8 &{\bf 173.1} & 206.6 \\\hline
\end{tabular}}
\end{table}

\subsection{Online Matching}
In this part, the performance of online algorithms (i.e., MA, MH, MED, and LA) in matching flexible customers is analyzed across different loading and generation scenarios.
\subsubsection{Scenario 1}
In this scenario the arriving customers have high criticality and short deadlines. The mean level of renewable generation varies and peaks during the middle of the epochs but is less than the arriving new load requests at all times (see Fig. \ref{fig.Load_Profile}). Hence, we can expect that waiting to serve the customers with delay will not be advantageous and can reduce the performance. Not surprisingly, we find that MA which matches on arrival achieves the highest social welfare. The average welfare achieved by each of the algorithms is shown in Table \ref{tab:Numerical_results}. While MA achieves the highest average welfare of $150.4$, the learning algorithms LA1 and LA2 are the second best at an average score of $130.5$ and $126.4$. 

\begin{figure}[h]
\centering{
\includegraphics[width=1\linewidth]{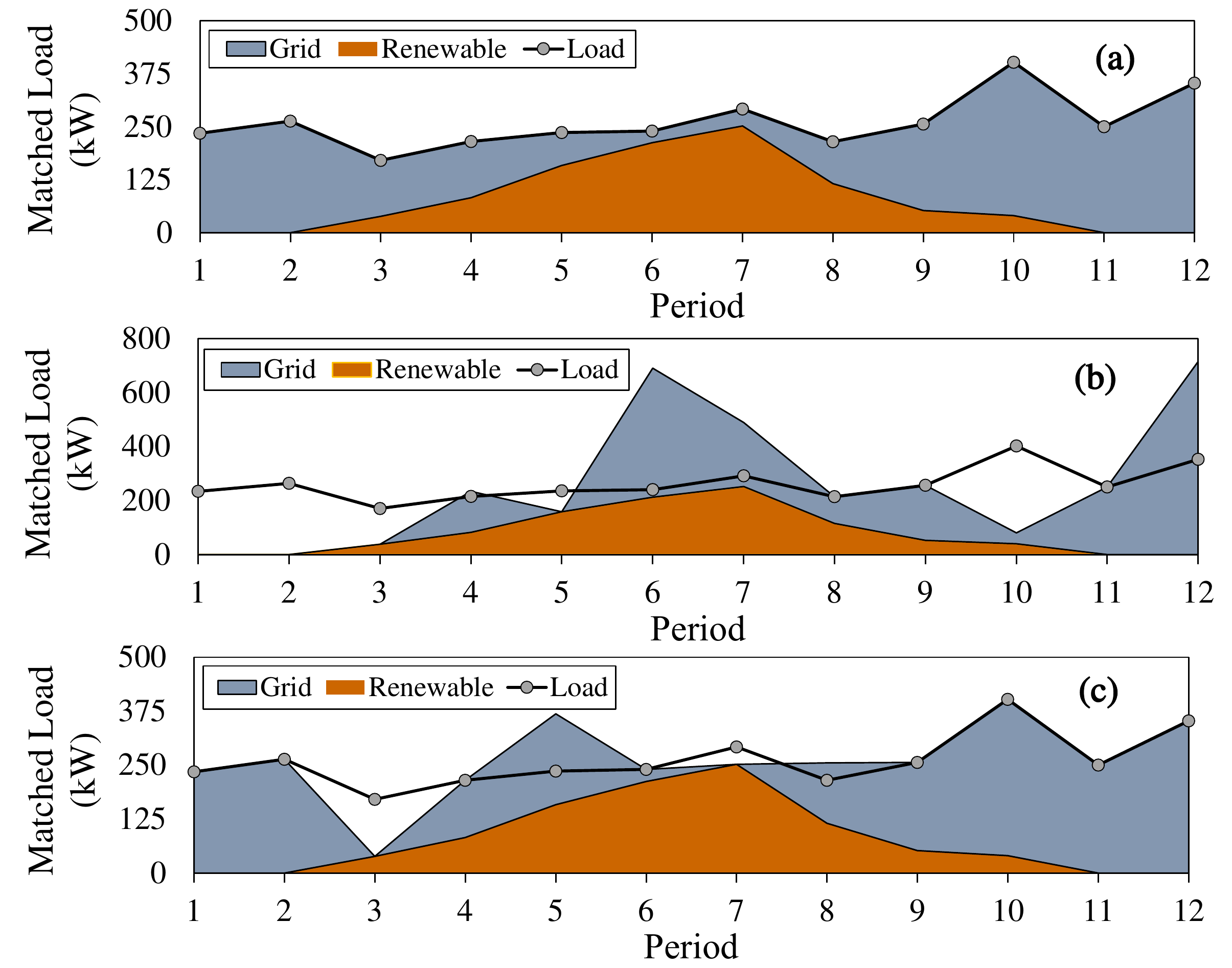}}
\caption{\small Matching on a representative epoch (Scenario 1): (a) MA, (b) MH, (c) LA1.}
\label{fig.Ex1_Results}
\end{figure}

The matching by MA, MH and LA1 on a representative epoch is shown in Fig. \ref{fig.Ex1_Results}. The figure shows how the learning algorithm matches after it learns. Here, we find that the learning algorithm was able to learn to match nearly on arrival. 



\subsubsection{Scenario 2}
\begin{figure}[h]
\centering{
\includegraphics[width=1\linewidth]{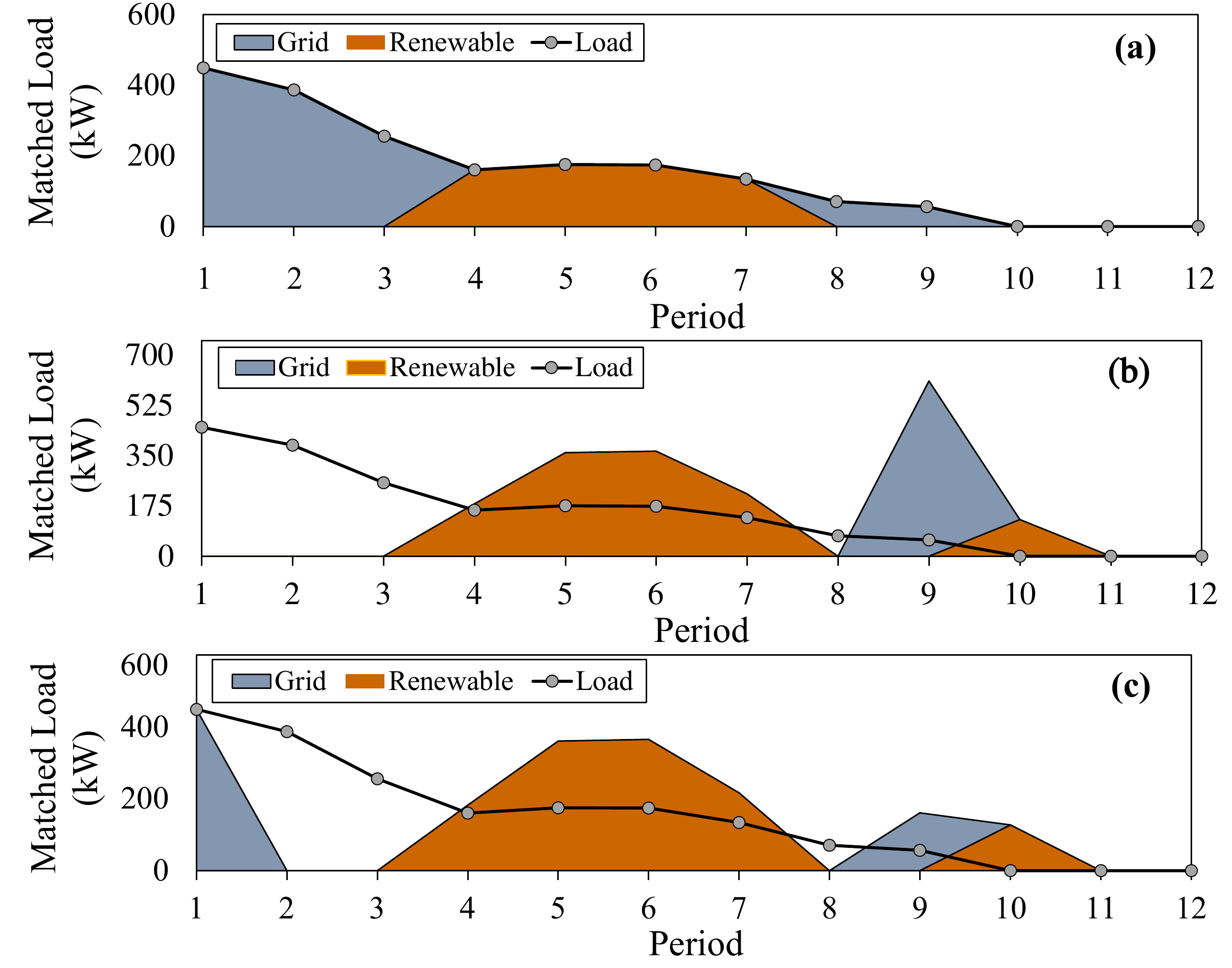}}
\caption{Matching on a representative epoch (Scenario 2): (a) MA, (b) MH, (c) LA1.}
\label{fig.Ex2_Results}
\end{figure}

In this scenario customers are less critical, and some customers which arrive during the middle of the epoch have a smaller deadline and some others have a larger deadline. The renewable generation is in excess during the middle of the epoch and the later part of the epoch while it falls severely short of the arriving load request at other times (see Fig. \ref{fig.Load_Profile}). Hence the MA algorithm which matches on arrival can under-utilize the renewable generation both at the middle and the end of the epoch while in contrast MH can maximally utilize the renewable generation at both these times because it waits till it finds sufficient renewable generation. This is evident in Fig. \ref{fig.Ex2_Results}, where matching by MA, MH and LA1 on a representative epoch is illustrated.  

The figure clearly shows that MH is able to leverage the customers' flexibility to match the renewable energy in excess of the arriving load request during the middle and the end of the epoch. As a result the majority of customers are matched to the renewable energy and the remaining critical ones are served by the grid. From the figure it is clear that the learning algorithm has learnt to delay the customers to utilize the renewable generation during the middle and end of the epoch. In addition, the learning algorithm does not delay all of the customers as MH does and matches some of the customers on arrival (see Fig \ref{fig.Ex2_Results}). The average welfare achieved by the algorithms for this scenario is shown in Table \ref{tab:Numerical_results}. We find that the learning algorithm LA1 achieves the highest average welfare of $137.4$, while MH achieves the second best average score of $136.7$ closely followed by LA2 at $134.7$.
\subsubsection{Scenario 3}
In this scenario, the difference from scenario $2$ is the one peak in renewable generation instead of two. The matching algorithm can still gain from allowing the customers who arrive earlier and with longer deadlines to wait in order to match to the renewable generation during the peak. Therefore, not surprisingly we find that matching on arrival is not the best strategy for this scenario. The average welfare achieved by each of the algorithms for this scenario is shown in Table \ref{tab:Numerical_results}. We find that the learning algorithms LA2 and LA1 achieve a high average score of $202.1$ and $188.5$ respectively.

\begin{figure}[h]
\centering{
\includegraphics[width=1\linewidth]{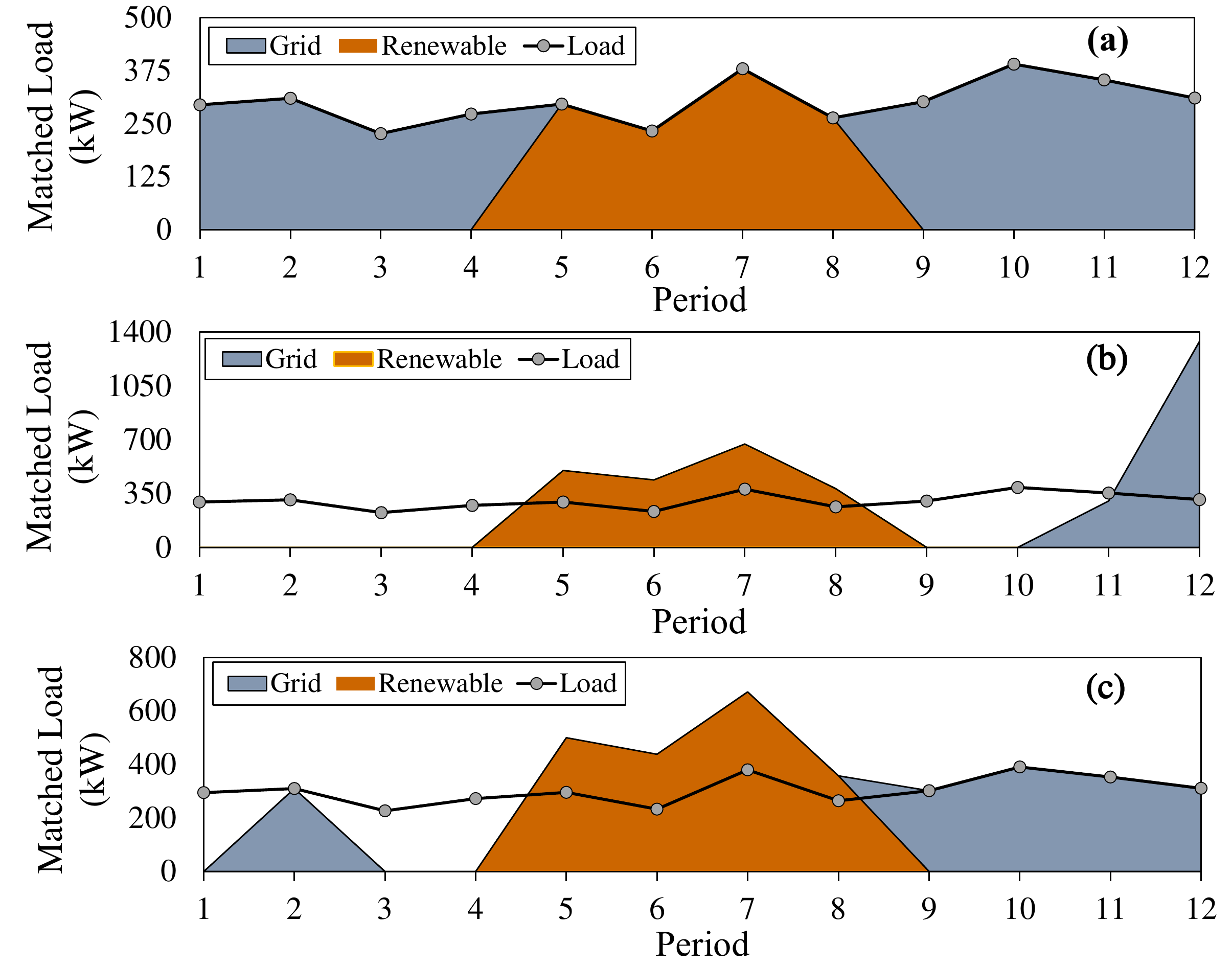}}
\caption{\small Matching on a representative epoch (Scenario 3): (a) MA, (b) MED, (c) LA1.}
\label{fig.Ex3_Results}
\end{figure}

\begin{figure}[h]
\centering{
\includegraphics[width=1\linewidth]{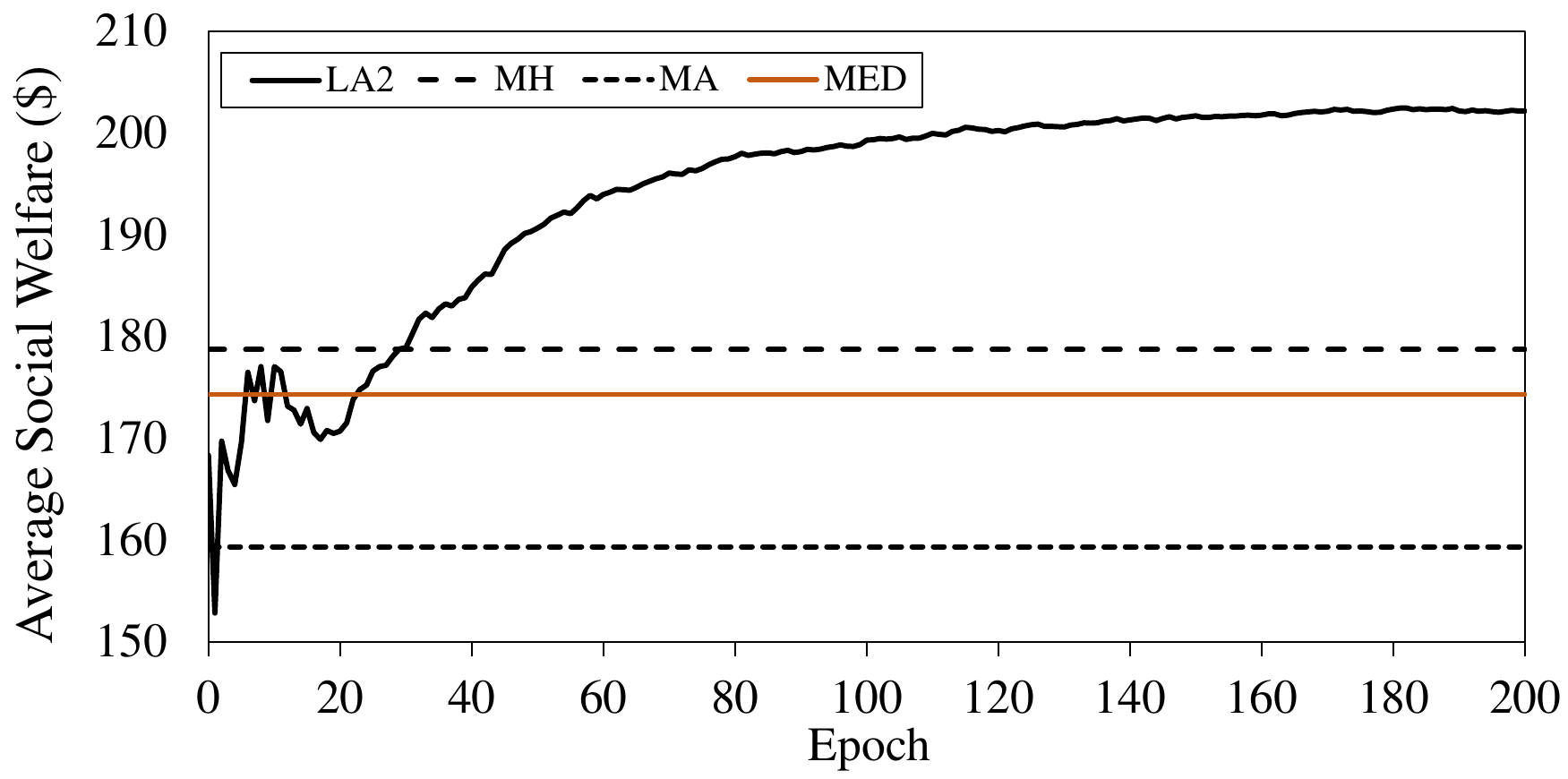}}
\caption{\small Average social welfare for LA2 (Scenario 3).}
\label{fig.Total_ave_Ex3}
\end{figure}

The matching by MA, MED and LA1 on a representative epoch is illustrated in Fig. \ref{fig.Ex3_Results}. The figure clearly shows that MA fails to leverage the customer's flexibility when compared to MED and LA1 to maximize the matching to renewable energy. This is because MA matches the customers who arrive during the earlier part of the epoch to the grid on their arrival itself and does not wait. 
While the learning algorithm has learnt to wait to avail future renewable generation, unlike MED, it matches some customers on arrival. As in scenario 2, this enables the learning algorithms to achieve a better score.  
Figure \ref{fig.Total_ave_Ex3} shows how the average social welfare improves for the learning algorithm LA2 with the number of epochs of data. The figure clearly highlights the importance of the role played by the learning component of the proposed matching policy. 

\subsubsection{Scenario 4}
In contrast to the previous three scenarios, here the mean level of new load request and renewable generation are uniform throughout and nearly equal. The average welfare achieved by each of the algorithms for this scenario is shown in Table \ref{tab:Numerical_results}.
We find that the learning algorithms LA2 and LA1 achieve the highest average welfare of $267.1$ and $266.9$.

\subsubsection{Scenario 5}
\begin{figure}[h]
\centering{
\includegraphics[width=1\linewidth]{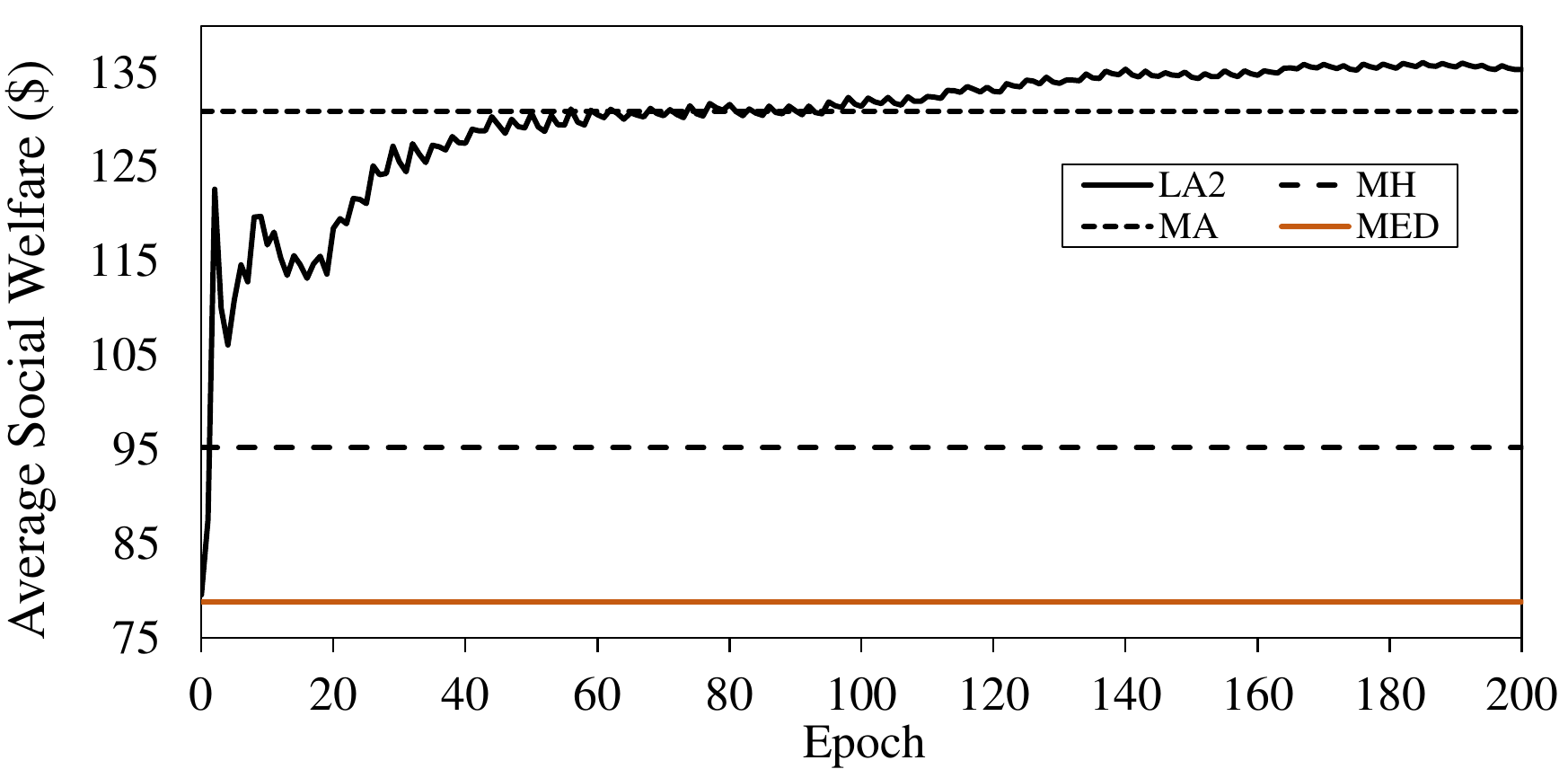}}
\caption{\small Average social welfare for LA2 (Scenario 5).}
\label{fig.SW_EX6}
\end{figure}

This scenario is a hybrid of scenarios $1, 2$ and $3$. On every $3k+1$th epoch, $3k+2$th epoch, $3(k+1)$th epoch, $k \in \{0,1,2,...\}$, the load and renewable generation are different and correspond to scenarios $1, 2$ and $3$, respectively. 
Unlike previous, the generation-consumption profiles for each subsequent epoch differ in this scenario. 
Figure \ref{fig.SW_EX6} shows how the average social welfare improves for LA2 with the number of epochs of data. 
Table \ref{tab:my-table} shows the social welfare achieved by the various algorithms on the $3k+1$th, $3k+2$th, and $3(k+1)$th epoch for $k = 54$ and $63$. We find that MA, MH and MED fluctuate from being the best on one epoch to the worst performing on the other epoch. 
This is expected from the observations we made on these algorithms under the sections for scenarios $1, 2$, and $3$. 
Most importantly, we find that the performance of learning algorithm LA2  does not vary as much as the other three algorithms and is consistently among the top two performing algorithms on each individual epoch. 
From Table \ref{tab:Numerical_results} we find that LA2 is the top performing algorithm overall with an average welfare of $135.3$ across all the epochs.





\begin{table}[h]
\caption{\small Social welfare in Scenario 5}
\label{tab:my-table}
\centering
\begin{tabular}{cccccc}
\hline
\multirow{2}{*}{\textbf{$k $}} & \multirow{2}{*}{\textbf{Epoch}} & \multicolumn{4}{c}{\textbf{Social Welfare (\$)}} \\ \cline{3-6} 
 &  & \textbf{MA} & \textbf{MH} & \textbf{MED} & \textbf{LA2} \\ \hline
 & \textbf{163} & 165.3 & 68.1 & 47.5 & 113 \\
\textbf{54} & \textbf{164} & 81.6 & 103 & 80.5 & 116.4 \\
 & \textbf{165} & 161.0 & 194 & 194.2 & 216.5 \\ \hline
 & \textbf{190} & 154.5 & 27 & 3.7 & 87.3 \\
\textbf{63} & \textbf{191} & 86.9 & 100.5 & 73.6 & 112.2 \\
 & \textbf{192} & 157.2 & 183.7 & 184.8 & 214.6 \\ \hline
\end{tabular}
\end{table}

\section{Conclusion}\label{Conls}
In this paper, a reinforcement learning model is proposed for dynamic matching of flexible customers and D-RES in power systems.
The proposed learning model is composed of a fixed rule-based function and a trainable component that can be directly trained by customer and D-RES data with no prior knowledge or expert supervision. 
The output from the proposed reinforcement learning model is the matching policy for active customers and D-RES owners, which seeks to maximize social welfare in the matching market while respecting the customers' servicing constraints.
Simulations are conducted on a test power system for different load and generation scenarios and the results show that the proposed reinforcement learning model outperforms the other standard online matching heuristics in terms of leaning an effective matching policy that achieves higher social welfare.
The results shows that the reinforcement learning model is much more effective than other online heuristics in determining matching policy across various scenarios that differ in terms of load and generation profiles and customers' servicing constraints.  
More specifically, it was found that the learning model is the best performing online algorithm across most of the scenarios, while the performance of other standard online heuristics varies intensely based on the supply availability and customers' servicing constraints.  

\bibliographystyle{IEEEtran}
\bibliography{Platform}

\end{document}